\documentclass{article}
\usepackage{amssymb}
\usepackage{amsmath}

\input{tcilatex}

\begin{document}

\title{On an additional realization of supersymmetry in orthopositronium
lifetime anomalies}
\author{B.M.Levin$^{1}$, V.I.Sokolov$^{2}$ \\
A.F.Ioffe Physical Technical Institute, \\
194021 St. Petersburg, Russia}
\maketitle

\begin{abstract}
Expansion of Standard Model for the quantitative description of the
orthopositronium lifetime anomalies (from QED to supersymmetric QED/SQED)
allows to formulate experimental tests of supervision of additional
realization of the supersymmetry in final state of the positron beta-decay
of the nuclei such as $^{22}$Na, $^{68}$Ga ($\Delta J^{\pi }=1^{+}$). The 
\textit{expermentum crucis} program is based on supervision of the
orthopositronium \textquotedblleft isotope anomaly\textquotedblright , on
the quantitative description of the \textquotedblleft $\lambda _{T}$ -
anomaly\textquotedblright\ and will allow to resolve the alternative as
results of the last Michigan work (2003).

\end{abstract}

No reliable observations of supersymmetry effects have thus far been
reported. Standard Model (SM) concepts suggest that this is due to the
presently achievable accelerator energies being not high enough, an obstacle
that may be overcome in the nearest future after the start-up in CERN of the
new accelerator, the Large Hadron Collider (LHC).

The manifestation of supersymmetry would be signaled by \textquotedblleft
disappearance\textquotedblright\ of the energy and momentum carried away by
the lightest of the supersymmetric particles [1] (the \textit{energy deficit}%
).

It is not inconceivable, however, that supersymmetry effects have already
manifested themselves on a conceptually different road which is not
associated with ultrahigh particle energies. We are going to show below that
an energy deficit can likewise be an experimental criterion of manifestation
of an \textit{additional realization of supersymmetry} beyond the present SM.

It is well known that double application of supertransformation, from the
fermion to boson and back to the fermion, transfers a particle to another
point in space. This is accepted as nothing more than a mathematical
attribute of supertransformations: \textquotedblleft \ldots the
anticommutator of two spinor generators\textit{\ }$Q$\textit{\ }is expressed
through the quantity of dimension\textit{\ }$m$, namely, through a 4-momentum%
\textit{\ }$p_{\mu }$\textit{, }the generator of the four-dimensional
displacement%
\begin{equation*}
\{Q,\overline{Q}\}\equiv Q\overline{Q}+\overline{Q}Q=-2p_{\mu }\gamma _{\mu
}\ ,
\end{equation*}%
where $\gamma _{\mu }$ are the Dirac's 4-matrices. The spinor transformation
may be conceived as if a square root of the displacement"\ [2, p.107]. But
it is in this \textit{mathematical fact} that \textit{new physics} probably
lies hidden.

The proposed program of \textit{additional measurements} would make use of
observation and quantitative description of the \textit{lifetime anomalies}
of orthopositronium (\textit{o-Ps}, $^{T}$\textit{Ps}), more specifically,
of the \textquotedblleft \textit{isotope anomaly}\textquotedblright\ in neon
in the \textquotedblleft \textit{resonance conditions}\textquotedblright\
(precision \textit{comparative} measurements yield the $1.85$ $\pm \ 0,1$
factor [3]) and of the \textquotedblleft $\lambda _{T}$-\textit{anomaly}%
\textquotedblright\ (precision \textit{absolute} measurements yield $\Delta
\lambda _{T}$/$\lambda _{T}$ $\simeq $ $+\ /0.19\div 0.14/\ \%$), see
references in [4]).

Progress reached in the quantitative description of the orthopositronium
lifetime anomalies [5] permits us to consider \textit{displacement} as a
structural element of \textit{a non-stationary long-range
non-Newtonian/non-Coulombian type for all physical interactions}, including
long-range (non-stationary!) baryon charge interaction in the final state of
positron\ $\beta $-decay of nuclei of the type of $^{22}$Na and $^{68}$Ga ($%
\Delta J^{\pi }$ $=1^{+}$). In so doing,\ $\beta ^{+}$-decay is considered
as \textit{a topological quantum transition} (TQT), and the final state of
the nuclei exists during the time$\ \tau _{\mu }$\ $\sim \ 2\cdot 10^{-6}$ s
against the background of a \textquotedblleft \textit{defect}%
\textquotedblright , a limited macroscopic volume of a crystal-like
space-time structure (the \textquotedblleft \textit{long-range atom}%
\textquotedblright ). Displacement plays here the part of the constant of a
three-dimensional cubic \textit{lattice}, but a 3D structure may be
conceived of as a discrete one-dimensional structure (\textit{Hamiltonian
cycle} over $N^{(3)}$ \textquotedblleft \textit{sites}\textquotedblright )
during the time $\tau _{\mu }$ [6]. This model permits one to calculate the
total number of sites in the \textquotedblleft \textit{atom}%
\textquotedblright\ (the number of \textit{steps} in the cycle), $N^{(3)}$ $%
=1.302\cdot 10^{19}$. The \textquotedblleft atom\textquotedblright\ can be
differentiated into a \textquotedblleft \textit{nucleus}\textquotedblright\
(the number of \textquotedblleft sites\textquotedblright\ $\overline{n}%
=5.2780\cdot 10^{4}$) and a \textquotedblleft \textit{shell}%
\textquotedblright . All sites in such an \textquotedblleft
atom\textquotedblright\ are identical in the full set of \textit{charges} of
all physical interactions while differing (in the \textquotedblleft
nucleus\textquotedblright\ and the \textquotedblleft shell\textquotedblright
) in their dynamic manifestations [5, 7].

The success of this model consists not only in that it is capable of a
quantitative description of the "isotope anomaly" (factor 2) and of the "$%
\lambda _{T}$-anomaly" of \textit{o-Ps}, but in its having expressed to
within $\sim \ 10^{-3}$ the double-valued ($\pm $) Planckian mass through
the fine structure constant$\ \alpha \ $[6]

$\pm \ M_{Pl}=\pm \sqrt{\frac{(\pm \hbar )\cdot (\pm c)}{G}}\simeq
N^{(3)}\cdot \lbrack (\pm m_{p})+(\pm m_{e})]\simeq \frac{2^{9/2}}{3\pi
^{2}\alpha ^{9}}\cdot \lbrack (\pm m_{p})+(\pm m_{e})]\simeq 2.179\cdot
10^{-5}$g.

The issue of the existence of new long-range forces was formulated in the
SM: theory has considered the consequences of a possible existence of
long-range interactions, of both abelian and non-abelian type [2, pp.
122-123], and on the experimental side, a high-precision search for a
fundamental spin long-range interaction has yielded a negative result [8].

The present SM offers a fairly loose treatment of the \textit{null energy
condition} (NEC) as \textit{absolute forbiddenness} of physical
manifestation of the negative energy (mass) sign. Note, however, that the
concept of negative energy (mass) density of the compensating field does not
involve any difficulties of a fundamental nature.

In TQT conditions, the argument of impossibility of a physical manifestation
of spacelike structures likewise loses its persuasiveness (\textit{causal
anomalies}).

A basic possibility of overcoming this forbiddenness is offered by the
concept of "... \textit{a complete relativity}, i.e., of equivalence of all
velocities (except for the light itself)", which allows the existence of "%
\textit{non-electromagnetic long-range interaction of bodies with a nonzero
average spin density}" [9] (cf. [8]).

Complete relativity (i.e. coexistence of \textit{locality} and \textit{%
non-locality}) is formulated in other terms (A. L. Zelmanov's \textit{method
of chronometric invariants}) in the theory of \textit{zero-space} and 
\textit{zero-particles} \textit{in a generalized space-time}, which was
developed independently as an expansion of the \textit{General Theory of
Relativity} [10, 11].

The orthopositronium anomalies have created a new situation. Removal of
these fundamental limitations in an analysis of the nature of the \textit{%
o-Ps} anomalies provides a justification for simultaneous physical
realization in the final state of $\beta ^{+}$-decay (or, treated in a
broader sense, in the final TQT state [7]) of positive Planckian mass as 
\textit{vacuum-like states of matter}/VSM [12] ($+\ M_{Pl}$: a VSM "\textit{%
microstructure}") and of the compensating "\textit{C-field}" (the "\textit{%
mirror Universe}", $-\ M_{Pl}$) [5-7].

All this can also be treated in the context of \textit{non-abelian-type}
long-range forces: "\ldots \textit{Yang-Mills theory with zero mass
obviously does not exist, because a zero mass field would be obvious; it
would come out of nuclei right away. So they \TEXTsymbol{<}meson physicists%
\TEXTsymbol{>} didn't take the case of zero mass and not investigate it
carefully}"\ [13] (cf. [2, pp. 122-123]).

At the \textit{University of Michigan} (Ann Arbor), the last measurement of
the \textit{o-Ps} annihilation rate has been carried out; the researchers
report now on the complete agreement between the experimental value, $%
\lambda _{T}$(\textit{exp}) $=7.0404(10)(8)$ $\mu $s$^{-1}$, and the value
calculated in the frame of QED, $\lambda _{T}$(\textit{theor}) $%
=7.039979(11) $ $\mu $s$^{-1}$ [4]. These measurements were performed by a
different technique, namely, a dc electric field of $\sim $7 kV/cm was
introducted into the measurement cell, and, therefore, the final conclusion
of the (new) Michigan group can hardly be considered unambiguous, because it
disregards the \textquotedblleft isotope\ anomaly\textquotedblright\ of 
\textit{o-Ps}. For this reason, the researchers could not include the 
\textit{additional} action of the electric field on the observed \textit{o-Ps%
} self-annihilation rate $\lambda _{T}$(\textit{exp}) [6], besides the
provisions they undertook to ensure complete \textit{o-Ps} thermalization.
The additional action of the electric field \textit{E} $\sim $7 kV/cm
oriented parallel to the force of gravity should suppress the excess $\Delta
\lambda _{T}/\lambda _{T}\simeq $ $(0.19\pm 0.14)\ \%$ over the calculated
value $\lambda _{T}$(\textit{theor}), which had been reported earlier by the
Michigan group and reffered to quantitatively as \textit{macroscopic quantum
effect} (the "$\lambda _{T}$-anomaly"\ ref. [5-7]).

This is why rejection [4] of the conclusions drawn from the earlier
high-precision $\lambda _{T}$ measurements does not appear unambiguous.

The uncertainty we are presently witnessing can be resolved only by
performing a \textit{program of additional measurements}.

Consider the scheme of a \textit{Gedanken experiment} for a measuring cell
filled with a gas (Fig. 1).

\FRAME{ftbhFU}{4.2194in}{1.9277in}{0pt}{\Qcb{Scheme and the \textit{result}
of a Gedanken experiment with\textit{\ }an electric field in the Earth
laboratory. The measuring cell is filled with gas. $\protect%
\overleftrightarrow{\mathit{E}}$ -- orientation and dc voltage of an
electric field; V -- is the value of the parameter to be measured.}}{\Qlb{%
fig1}}{fig_1.jpg}{\special{language "Scientific Word";type
"GRAPHIC";maintain-aspect-ratio TRUE;display "USEDEF";valid_file "F";width
4.2194in;height 1.9277in;depth 0pt;original-width 5.2399in;original-height
2.3748in;cropleft "0";croptop "1";cropright "1";cropbottom "0";filename
'Fig_1.JPG';file-properties "XNPEU";}} \ \ 

Could one substantiate a program of comparative measurements which would
yield as a final result the \textit{doubling} (factor 2) of the parameter V
to be measured with the external dc electric field orientation changed from 
\textit{horizontal} to \textit{vertical}? This would be certainly impossible
within the SM. An analysis of the \textit{o-Ps} anomalies within the concept
of spontaneously broken complete relativity [5-7] opens up such a
possibility; indeed, restoration of the symmetry under discussion
\textquotedblleft \textit{should be accompanied }\TEXTsymbol{<}\ldots 
\TEXTsymbol{>} \textit{by doubling of the space-time dimension}%
\textquotedblright\ [9].

The uniqueness of orthopositronium dynamics (\textit{virtual} single-quantum
(!) annihilation, CP-invariance) make it an intriguing probe to double the
space-time (see [5]).

Unlike gravity and electricity, the new long-range interactions become
manifest only in transient (non-stationary) conditions, as a \textit{%
generalized }\textquotedblleft \textit{displacement current}%
\textquotedblright\ of all physical interactions (with double-valued
\textquotedblleft charges\textquotedblright , including the masses $\pm \
m_{p}$ and $\pm \ m_{e}$) and have the limited (macroscopic) \textit{radius
of action} $R_{\mu }$ $\sim \ 6\cdot 10^{4}$cm.

Consider in this connection again the standard experimental technique used
to measure positron/orthopositronium annihilation lifetime spectra.

Figure 2 presents a block diagram of a fast-slow lifetime spectrometer of
delayed $\gamma _{n}$-$\gamma _{a}\ $coincidences.

Recording of real coincidences (in the start-stop arrangement) with a time
resolution of $1.7\cdot 10^{-9}$ s [3] between the signal produced by a
nuclear $\gamma _{n}$-quantum of energy $\simeq 1.28$ MeV (\textquotedblleft
start\textquotedblright ) with the signal generated by the detected $\gamma
_{a}$ annihilation quantum of energy $\simeq $ ($0.34\ \div 0,51$) MeV
(\textquotedblleft stop\textquotedblright ) (corresponding, accordingly, to 3%
$\gamma $- and 2$\gamma $ annihilation) is accompanied by the energy
(amplitude) discrimination in the slow (\textquotedblleft
side\textquotedblright ) coincidence channels (with a resolution $\delta
\tau $ $\sim \ 10^{-6}$ s) between the corresponding signals from the
last-but-one dynodes of the lifetime PM tubes, an approach that cuts
efficiently \textit{random} coincidence noise.

After subtraction of the random coincidence background, the positron
annihilation lifetime spectra of inert gases would represent the sums of
exponentials with characteristic annihilation rate constants $\lambda _{i}$%
\begin{equation*}
N(t)=\overset{i\ =\ 2}{\underset{i\ =\ 0}{\dsum }}I_{i}\cdot e^{-\lambda
_{i}\ t},
\end{equation*}%
where $\lambda _{0}$ and $I_{0}$ are, respectively, the rate and intensity
of the \textit{two-quantum} annihilation of the parapositronium component (%
\textit{p-Ps,}$^{S}$\textit{Ps}), $\lambda _{1}$ and $I_{1}$ are the
component of two-quantum annihilation of the quasi-free positrons that have
not formed positronium (with \textit{peculiarity}, so-called "\textit{%
shoulder}" [5]), and $\lambda _{2}$, $I_{2}$ are those of \textit{%
three-quantum annihilation} of the orthopositronium component.

It is known that single-quantum \textit{o-Ps} annihilation is forbidden in
QED by the momentum conservation law. There is no such limitation in
supersymmetric QED (N = 2 SQED) because of \textit{total degeneracy of the
para- and ortho-superpositronium} forming by virtual single-quantum \textit{%
o-Ps} annihilation into the \textquotedblleft mirror
Universe\textquotedblright\ as a result of \textit{doubling} of the
space-time dimension in the final TQT state [5].

Experimental bounds accumulated in the two decades of intense studies of the
orthopositronium problem lead one to the conclusion that the \textit{%
additional single-quantum mode} of orthopositronium annihilation involves
not a photon but rather a \textit{notoph} ($\gamma ^{o}$ is a zero-mass,
zero-helicity particle which is complementary in properties to the photon)
[14] and two \textit{mirror photons} $\gamma ^{\prime }$ with a negative
total energy of $3.6\cdot 10^{-4}$ eV:%
\begin{equation*}
^{T}Ps\backslash ^{T}Ps^{\prime }(^{s}Ps^{\prime })\longrightarrow \gamma
^{o}\backslash 2\gamma ^{\prime }.
\end{equation*}

This was how the broadening of the framework in which the nature of the 
\textit{o-Ps} anomalies could be analyzed (from QED to SQED) and the
phenomenology of the mechanism of energy and momentum deficit compensation
in a single-quantum mode were first formulated [15].

Treated from the SM standpoint, however, detection of a quantum of energy $%
1.022$ MeV in the \textquotedblleft stop\textquotedblright\ channel of the
fast-slow coincidences is forbidden (see the \textquotedblleft
lower\textquotedblright\ and \textquotedblleft upper\textquotedblright\
detection thresholds of $\simeq $ $(0.34\div 0.51)$ MeV, respectively, in
Fig. 2).

We are now coming back to the principal question of \textit{how the
additional realization of supersymmetry would be established in the
experiment}.

Detection of a \textit{single-notoph o-Ps annihilation mode} should also be
accompanied by observation of an \textit{energy deficit} in the
\textquotedblleft stop\textquotedblright\ channel of the lifetime
spectrometer: indeed, \textit{single-notoph annihilation} is identified in
the scintillator by the Compton-scattered electron $e$, which is bound in
the \textit{long-range atom} \textquotedblleft shell\textquotedblright\ in a
\textquotedblleft pair\textquotedblright\ $e\overline{e}$ with the
\textquotedblleft electronic hole\textquotedblright\ $\overline{e}$ (\textit{%
negative mass}) in the \textquotedblleft C-field\textquotedblright
/\textquotedblright mirror-Universe\textquotedblright\ structure. Half of
the notoph energy, $\simeq $ $0.51$ MeV , is transferred to the $e\ $hole ($%
\overline{e}$) and, thus, \textquotedblleft disappears\textquotedblright\
(\textquotedblleft \textit{anti-Compton scattering}\textquotedblright ). As
a result, the additional single-notoph mode is detected by the lifetime
spectrometer in the \textquotedblleft stop\textquotedblright\ channel by
Compton scattering of an electron $e$ of energy $\leqslant $ $0.51$ MeV.%
\FRAME{ftbhFU}{4.1087in}{3.6288in}{0pt}{\Qcb{Block-diagram of the lifetime
spectrometer (fast-slow $\protect\gamma _{n}$-$\protect\gamma _{a}$
coincidences). \textbf{ID}\ -- integral discriminator (excludes $\protect%
\gamma _{a}$ detection in the \textquotedblleft start\textquotedblright\
channel); \textbf{DD} -- Differential Discriminator (restricts $\protect%
\gamma _{n}$ detection in the "stop" channel); \textbf{SCM} -- Slow
Coincidence Module; \textbf{TAC} -- Time-to-Amplitude Converter ($\Delta
t\longrightarrow $A); \textbf{MPHA} -- Multichannel Pulse-Height Analyzer.}}{%
\Qlb{fig2}}{fig_2.jpg}{\special{language "Scientific Word";type
"GRAPHIC";maintain-aspect-ratio TRUE;display "USEDEF";valid_file "F";width
4.1087in;height 3.6288in;depth 0pt;original-width 2.0401in;original-height
1.8005in;cropleft "0";croptop "1";cropright "1";cropbottom "0";filename
'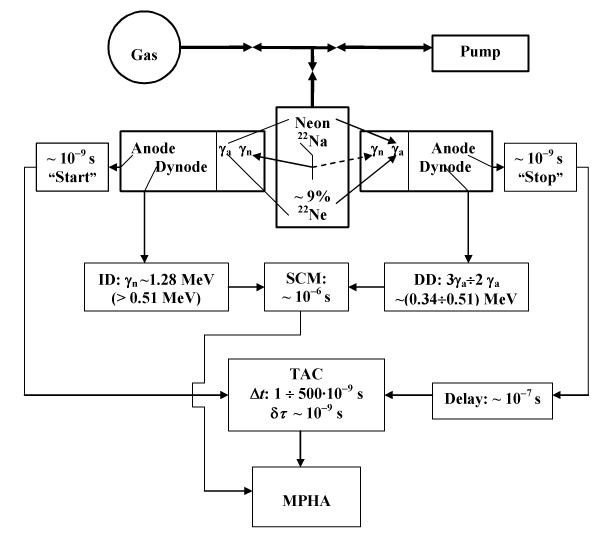';file-properties "XNPEU";}}

The experiment [3, 4] is in agreement with the phenomenology proposed for
quantitative description of the \textit{o-Ps} anomalies provided we assume
that the additional single-notoph annihilation mode contributes to the 
\textit{instantaneous coincidence peak} [5]. This means that one half of the
intensity of the long-lived lifetime spectral component obtained under
\textquotedblleft resonance conditions\textquotedblright\ for neon of
natural isotope abundance ($I_{2}$) transfers to the $t$ $\sim \ 0$ region.
An electric field of $\sim \ 7\ kV/cm$ applied parallel to the force of
gravity should suppress the additional mode and double the orthopositronium
component ($2I_{2}$). Accordingly, in the Michigan experiment (non-resonance
conditions) an electric field oriented along the force of gravity would
bring about complete agreement between $\lambda _{T}($\textit{exp}$)$ with
the QED-calculated value $\lambda _{T}($\textit{theor}$)\ $[4]; and the
disagreement of about$\ \Delta \lambda _{T}/\lambda _{T}\simeq (0.19\div
0.14)\ \%$ found previously (in experiments without electric field) should
again appear after the action of the electric field has been neutralized (by
applying it perpendicular to the force of gravity) [6].

The term \textquotedblleft \textit{anti-Compton scattering}%
\textquotedblright\ has been borrowed from Ref. [16]; it appears appropriate
to cite here an excerpt from the abstract of this paper written by a classic
of the theory of relativity:

"\textit{The purpose of this paper is to answer the following question in
terms of concepts of classical relativistic mechanics: How is Compton
scattering altered if we replace the photon by a particle of zero rest mass
and negative energy, and apply the conservation of 4-momentum? \TEXTsymbol{<}%
\ldots \TEXTsymbol{>} Since particles with negative energies are not
accepted in modern physics, it is perhaps best to regard this work as a
kinematical exercise in Minkowskian geometry, worth recording because the
results are not obvious}".

Observation of orthopositronium anomalies gives one physical grounds to
broaden the present-day SM. It appears now appropriate to analyze the
\textquotedblleft \textit{anti-Compton scattering}\textquotedblright\ in
connection with the detection of notoph in the proposed \textit{program of
additional measurements}, which aim at proving the existence of a connection
between the gravity and electromagnetism [5-7].

We may add that the concept of the supersymmetric version of a spin-$\frac{1%
}{2}$ \textit{quasi-particle} and \textit{hole} as \textit{supersymmetric
partners} [17].

\FRAME{fbhFU}{4.0274in}{2.0271in}{0pt}{\Qcb{Scheme of additional
measurements: is there a connection between gravity and electromagnetism? $%
I_{2}$ -- intensity of the orthopositronium ($^{T}Ps$) lifetime component
(with $^{22}$Na as a source of the positrons in a cell filled with gas) for 
\textit{neon }of natural isotope abundance (in the "resonance conditions": $%
\sim \ 9\ \%$ $^{22}$Ne) placed in a dc electric field $\sim $ $7\ kV/cm$ 
\textit{perpendicular} to the force of gravity; $2I_{2}$ -- same in an
electric field $\sim \ 7\ kV/cm$ \textit{parallel} to the force of gravity (%
\textit{doubling}).}}{\Qlb{fig3}}{fig_3.jpg}{\special{language "Scientific
Word";type "GRAPHIC";maintain-aspect-ratio TRUE;display "USEDEF";valid_file
"F";width 4.0274in;height 2.0271in;depth 0pt;original-width
5.0004in;original-height 2.5002in;cropleft "0";croptop "1";cropright
"1";cropbottom "0";filename 'Fig_3.JPG';file-properties "XNPEU";}}

To sum up: one should carry out additional measurements, because the result 
\textit{inconceivable} in the frame of the SM becomes an \textit{expected}
result in the program of \textit{experimentum crucis} (Fig. 3).

A positive result of this crucial experiment would mean the birth of \textit{%
new physics} that would be \textit{complementary} to the SM; this physics
would restore also spontaneous violation of the \textit{chiral symmetry},
which would be essential in elucidating the origin not only of the mass of
matter [18] but of the \textit{dark matter in the Universe} as well (see
[18]) [19].

\bigskip

$^{1}$E-mail: bormikhlev@mail.ioffe.ru\ , bormikhlev@mail.ru

$^{2}$E-mail: v.sokolov@mail.ioffe.ru

\end{document}